\documentclass[preprintnumbers,showpacs,twocolumn,secnumarabic,amssymb,amsmath,amsfonts,aps,pra]{revtex4}

\usepackage{graphicx}

\usepackage{amsmath}
\begin{document}

\newcommand{\non}{\nonumber}
\newcommand{\be}{\begin{equation}}
\newcommand{\ee}{\end{equation}}
\newcommand{\bq}{\begin{eqnarray}}
\newcommand{\eq}{\end{eqnarray}}
\newcommand{\lps}{\langle}
\newcommand{\rps}{\rangle}
\newcommand{\bs}{\hat a}
\newcommand{\bsd}{\hat a^\dagger}

\title{Quantum localization and bound state formation in Bose-Einstein condensates}

\date{November 18, 2010}

\author{Roberto Franzosi}
\author{Salvatore M. Giampaolo}
\author{Fabrizio Illuminati}
\affiliation{Dipartimento di Matematica e Informatica,
Universit\`a degli Studi di Salerno, Via Ponte don Melillo,
I-84084 Fisciano (SA), Italy, CNISM Unit\`a di Salerno, and INFN Sezione di Napoli,
Gruppo collegato di Salerno, I-84084 Fisciano (SA), Italy}

\begin{abstract}
We discuss the possibility of exponential quantum localization in systems
of ultracold bosonic atoms with repulsive interactions in open optical
lattices without disorder. We show that exponential localization occurs
in the maximally excited state of the lowest energy band. We establish the
conditions under which the presence of the upper energy bands can be neglected,
determine the successive stages and the quantum phase boundaries at which localization
occurs, and discuss schemes to detect it experimentally by visibility measurements.
The discussed mechanism is a particular type of quantum localization that is intuitively
understood in terms of the interplay between nonlinearity and a bounded energy spectrum.
\end{abstract}

\maketitle

\section{Introduction}

The phenomenon of Anderson localization in disordered quantum systems \cite{Review85}
was originally discovered in the context of the study of electrons in a crystal
with imperfections \cite{Anderson}. In fact, it is
much more general \cite{Kramer} and has been observed in a variety of systems,
including light waves in random media \cite{Albada,Wiersma}. Despite remarkable
efforts, Anderson localization has not been observed
directly in crystals, owing to the high electron-electron and electron-phonon
interactions.
It has finally been observed in non-interacting Bose-Einstein
condensates in one-dimensional quasi-periodic optical lattices \cite{Firenze},
that feature a crossover between extended and exponentially localized
states, as in the case of purely random disorder in higher dimensions; moreover,
the effects of random disorder in optical lattices can also be simulated manipulating
the interactions in multi-species mixtures \cite{QuantumEmulsions}.
These achievements are due to the unprecedented
degree of control over the system physical parameters, in particular the
vanishing of the interaction strength, that ultracold atoms offer.

Indeed, ultracold degenerate gases in optical lattices provide an unprecedented toolbox for the experimental
realization of what were once just toy models sketching the key features of complex condensed matter
systems. One prominent example is the Bose-Hubbard model \cite{Haldane_PLA_80_280,Fisher_PRB_40_546},
that was originally introduced as a variant of the
better known Hubbard model  and whose properties were later discussed at length in the context of the description of superfluid $^4$He trapped in porous media.
The suggested realization in optical lattices loaded with ultracold bosonic atoms \cite{Jaksch_PRL_81_3108} was soon achieved in a spectacular breakthrough experiment
\cite{Greiner_Nature_415_39}. Driven by this brilliant result, a growing number of investigations has
focused on the possibility to use optical lattices to realize various phenomena of considerable interest
in condensed matter physics \cite{BlochReviews,IlluminatiAlbus}. Amongst these, in the last years much attention has been devoted to the study of localized quantum phases in many-body systems.
For instance, it has been showed that it is possible to use boundary dissipation \cite{Franzosi} or the control of the sign of the local interactions, exploiting Feshbach resonances, to switch from the repulsive Hubbard model to the attractive one, whose ground state may feature a collapse of all the atoms of the system into a single site of the lattice \cite{Dorignac,Jack,Buonsante}. The transition to collapse is essentially due to the combination of the nonlinear dependence of the local Hamiltonian on the site occupation that makes energetically favorable those states that are characterized by a concentration of all atoms in a single site.

In the present work we describe a route to quantum localization in many-boson systems with repulsive
interactions, that has one important feature in common with the transition to collapse discussed in Refs.\cite{Dorignac,Jack,Buonsante}. The basic idea is to consider the maximally excited state in the lowest
energy band of an interacting system in a non-translationally invariant lattice. Within the Bose-Hubbard framework this is just the eigenstate with highest energy of an Hamiltonian $H$ with repulsive interactions. This state is then the eigenstate with lowest energy (i.e. the ground state) of a new Hamiltonian 
$H^{'}$ equal to minus the original Hamiltonian: $H^{'} = - H$. The rotated Hamiltonian has
attractive interactions (instead of repulsive interactions) and a negative tunneling amplitude. 
However, this can be turned positive again by a $\pi$-phase shift on every other lattice site.
If one can show the occurrence of exponential localization in the highest excited state of the
repulsive Hamiltonian, this phenomenon is then completely equivalent to the collapse
in the ground state of the corresponding attractive model. In the latter case, upon increasing the
intensity of the attraction, the particles form a bound state, with increasing mass, which appears to be localized if the correlation length becomes smaller than the lattice spacing.
The two mechanisms differ in that the former is not realized as an on-site collapse
in the ground state of a system with attractive interactions, but rather as a proper exponential
localization in the maximally excited state of the lowest energy band in systems with repulsive interactions. It is thus a mechanism that is solely due to the interplay between nonlinearity and an energy spectrum
bounded from above. The latter in turn is a fundamental feature associated to the
fact that the Bose-Hubbard Hamiltonian preserves the total particle number.

In the following we will investigate the properties of
the maximally excited state of the one-dimensional repulsive Bose-Hubbard model 
defined on open lattice chains. Such analysis, besides resorting to exact diagonalization for systems
of very small size, will be carried out for larger systems using numerical solutions obtained with a
controllable recursive algorithm as well as a semi-classical
approach when the size of the problem makes the numerical computation impractical.
We will show that, depending on the physical parameters
of the system, the maximally excited state of the repulsive Bose-Hubbard Hamiltonian defined
on an open chain features three different phases of which the first one,
associated to small values of the local repulsion, is characterized by a relative atomic population
spread on all sites of the lattice. At intermediate values of the on-site interaction there
occurs a second phase in which a macroscopic fraction of the atoms begins to localize in a single
site while the remainder of the atomic population is still spread over the lattice.
Finally, we will show that for large values of the on-site repulsion the maximally excited state is characterized by an exponential localization in the center of the lattice and we will investigate the decay rate both numerically and analytically.
We will then determine the physical conditions such that the overlap of the
maximally excited state of the lowest energy band with the lowest states of the upper bands
can be neglected, and we will discuss how to detect experimentally the three different behaviors
by measurements of the visibility.

\section{Model and methods}

Let us consider a system of $N$ ultracold atoms with repulsive on-site interactions
described by a Bose-Hubbard model on a one-dimensional lattice of $M$ sites:
\begin{equation}
H=  \frac{U}{2} \sum^{d}_{ j=-d} \hat n_j(\hat n_{j}-1)  
-  T \sum^{d-1}_{j=-d}
\left (\hat a^\dagger_{j} \hat a_{ j+1} + \textrm{h.c.} \right ) \, .
\label{BHH}
\end{equation}

One needs to consider open chains to look, even in principle, for the possibility
of true localization. Indeed, in a translationally-invariant geometry the atoms
would be unable to localize on a definite site. Namely, even in the presence
of strong repulsive on-site interactions, the maximally excited state would be essentially
a Schr\"odinger-cat state, i.e. a superposition of localized states characterized by
a flat distribution of the atomic density over the entire lattice \cite{Buonsante}.
In Eq.(\ref{BHH}) $d=(M-1)/2$, $\hat a_{j}$ ($\hat a^\dagger_{j}$) are
the bosonic annihilation (creation) operators on the $j$-th site,
$\hat n_{ j}=\hat a^{\dagger}_{ j} \hat a_{ j}$ are the occupation number operators,
$U>0$ is the strength of the repulsive nonlinear on-site interaction, and $T$
is the hopping amplitude between neighboring sites.

\begin{figure}
{\includegraphics[width=8cm]{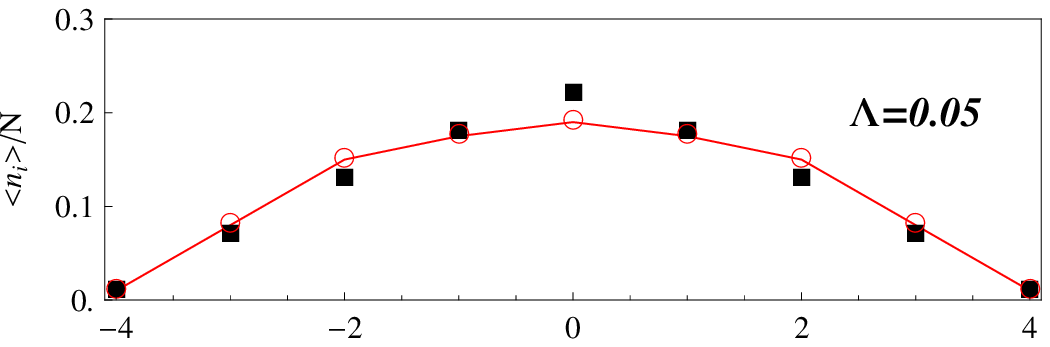}\vspace{-.18cm}
\includegraphics[width=8cm]{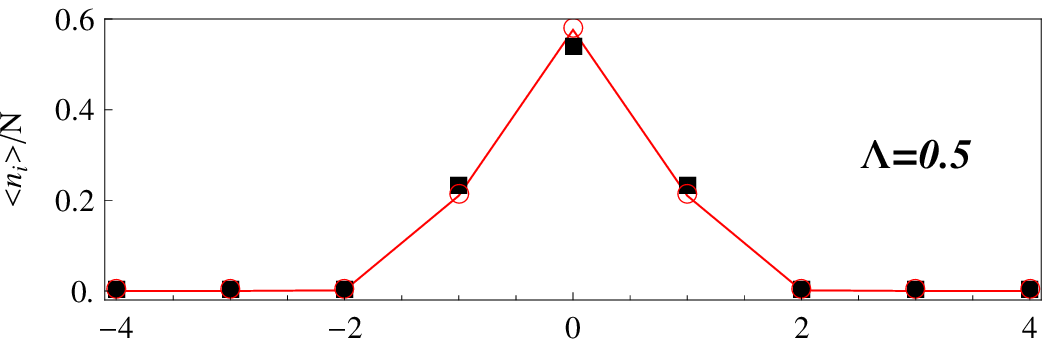}\vspace{-.18cm}
\includegraphics[width=8cm]{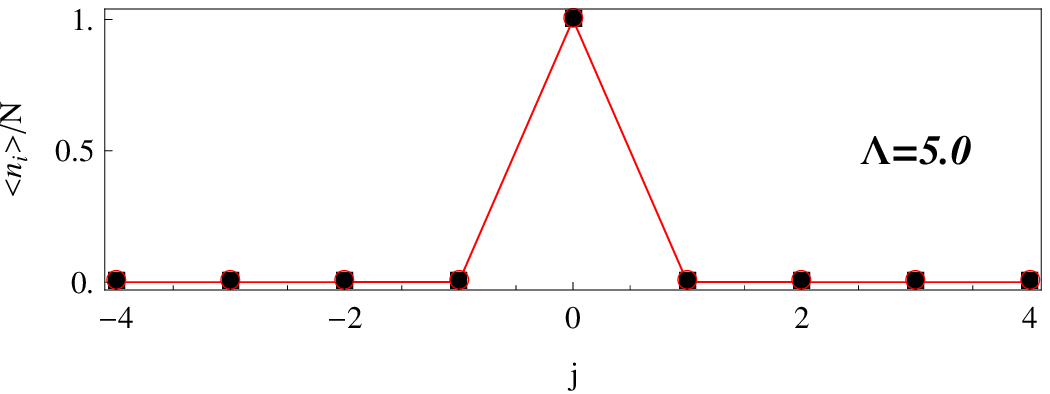}}
\caption{(Color online) The atomic density profile $\langle n_j \rangle/N$ in the maximally excited state
of Hamiltonian (\ref{BHH}) obtained by exact diagonalization (full black squares)
and by the semi-classical variational approximation (empty red
circles linked by solid red line), for different values of the dimensionless energy ratio
$\Lambda$ in an open chain of $M=9$ sites and $N=9$ atoms. All quantities being plotted
are dimensionless.}
\label{approx}
\end{figure}

In order to determine an optimized analytic approximation to the maximally excited state
of the Hamiltonian Eq.(\ref{BHH}) on a finite open chain we proceed by a dynamical variational
method and compare results with the ones obtained by exact diagonalization.
We follow the route adopted for the corresponding attractive model
\cite{Jack,Buonsante}, introducing a macroscopic trial state of the form
$|\tilde{\phi}\rangle=e^{i \varphi} |\phi\rangle$ where $\varphi$ is a time-dependent
phase and $|\phi\rangle$ is a coherent state of the form
\begin{equation}
 | \phi \rangle = \frac{1}{\sqrt{N!}}
\left( \sum^d_{j = -d} \phi_j \bsd_j \right)^N
|\Omega \rangle \, .
\label{sumcs}
\end{equation}
Here $|\Omega \rangle$ is the vacuum state and the coherent-state constants, $\phi_j \in \mathbb{C}$
for $j=1,\dots,M$, must satisfy the normalization condition $\sum^M_{j = 1} |\phi_j|^2=1$.
The complex quantities $\phi_j$ describe the on-site bosonic states by
the on-site population $|\phi_j|^2$ and the  macroscopic local phase $\arg{\phi_j}$.
The request that the trial state satisfies the Schr\"odinger equation on the average,
$\langle \tilde{\phi}| i \partial_t -H| \tilde{\phi} \rangle=0$, identifies the time
derivative $\dot{\varphi}$ with an effective Lagrangian for the dynamical variables
$\phi_j$ and the corresponding effective Hamiltonian \cite{Amico}:
\begin{equation}
{\cal{H}} =\frac{U}{2} N(N-1) \sum^d_{ j=-d} |\phi_j|^4
- T N \sum^{d-1}_{j=d}
\left (\phi^*_j \phi_{j+1} + cc \right )
\label{HS}
\end{equation}
Maximizing the latter with respect to the
variables $\phi_j$ under the normalization constraint, one obtains a semi-classical variational
approximation to the maximally excited energy eigenstate of the system.
The first term of Eq.(\ref{HS}), i.e. the on-site interaction term,
does not depend on the phases of the on-site variables $\phi_j$,
while the hopping term, at arbitrarily fixed values of $|\phi_j|$ and $|\phi_{j+1}|$ ,
is maximized by a phase difference $\pm \pi$.
Therefore, except for an irrelevant
global phase factor, the values $\phi_j$ associated to the maximally excited state can be
assumed to be real quantities with alternating signs. Defining $x_j=|\phi_j|$, and taking into
account the property of invariance under mirror reflection ($x_j=x_{-j}$) with respect to
the center of the finite chain, Eq.(\ref{HS}) can be recast in the equivalent forms
\begin{eqnarray}
\frac{\cal{H}}{N} & = & \frac{U}{2}  (N-1) \left( x_0^4 +2 \sum^d_{ j=1}  x_j^4 \right)
+ 2 T \sum^{d}_{j=1}  x_j  x_{j-1} \; , \nonumber \\
{\cal{H}} & = & \Lambda  \left( x_0^4 +2 \sum^d_{ j=1}  x_j^4 \right)
+ \sum^{d}_{j=1}  x_j  x_{j-1} \, .
\label{HS1}
\end{eqnarray}
Introducing the dimensionless ratio of the interaction to
kinetic energy scales: $\Lambda = U (N-1)/(8 T)$, one can solve
the problem for different values of $\Lambda$
and maximize Eq.(\ref{HS1}) using the hyperspherical representation 
of the variables $x_j$ to enforce automatically their exact normalization,
and can then compares the atomic distribution densities so obtained with the ones
provided by exact diagonalization for small samples.
The latter in turn can be performed very efficiently with the help
of augmented recursive Lanczos algorithms \cite{Andreozzi}.
The result of this comparison is reported in Fig.(\ref{approx}).

Fig.(\ref{approx}) shows that the semi-classical solution provides an excellent approximation
to the maximally excited state that becomes more and more accurate with increasing strength
of the interaction and of the localization of the atoms at the center of the lattice, and can
thus be extended to systems of much larger size that cannot be investigated by exact diagonalization.
More important, Fig.(\ref{approx}) shows that as the $\Lambda$ parameter varies the atomic
density profile in the maximally excited state crosses three different phases.
The first one, associated to small values of $\Lambda$, is characterized by the absence of
localization; the third one, associated to very large values of $\Lambda$, corresponds to
a complete concentration of all the atoms of the system in the center of the lattice;
finally, the second one, associated to intermediate values of $\Lambda$, corresponds to the
onset of the localization of a significant fraction of the atomic population in the
central site, while the distribution of the remainder of the atomic population over the
entire lattice stays finite.
\begin{figure}[t]
\includegraphics[width=7cm]{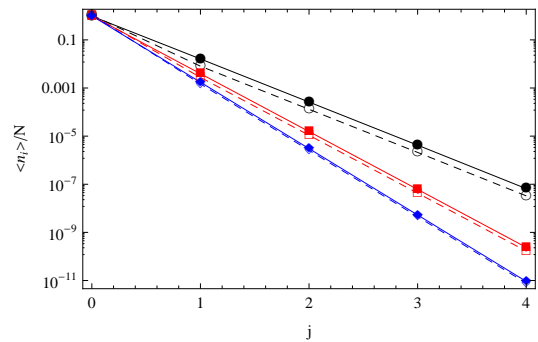}
\caption{(Color online) Atomic density distribution (in logarithmic scale)
associated to the maximally excited energy eigenstate in the
localized phase obtained by exact diagonalization
(solid lines and full symbols) and by the semi-classical variational
approximation (dashed lines and empty symbols) for a chain of $M=9$ sites
and $N=9$ atoms. From top to bottom: Black circles and lines ($\Lambda = 1.0$);
red squares and lines ($\Lambda = 2.0$); blue diamonds and lines ($\Lambda = 3.0$).
All quantities being plotted are dimensionless.}
\label{breather}
\end{figure}

\section{Exponential localization and visibility}
We now investigate in detail the exponential nature of the localization in
the maximally excited state. The presence of an exponential decay is reported
(in logarithmic scale) in Fig.(\ref{breather}). It shows the behavior of the
relative occupation $\langle n_i\rangle/N$ as a function of the distance from the center of the lattice.
The occurrence of an exponential localization allows to introduce a simple and
effective method to obtain an excellent analytical approximate solution the problem of the
maximization of the effective Hamiltonian. We introduce a dimensionless parameter $\chi$ and assume
$|\phi_j| = \chi^j \; \forall j \in [1,d]$, with $|\phi_0| = 1 - 2 \chi^2 (1-\chi^{2 d})/(1-\chi^2)$ in order to satisfy normalization. Within this setting the maximization of the effective variational Hamiltonian can be performed analytically. Exploiting the condition
$\chi \ll 1$, necessary to have a state localized in the central site, one finds the approximate
analytical solution $\chi = \Lambda / 8$.
\begin{figure}[t]
\includegraphics[width=7cm]{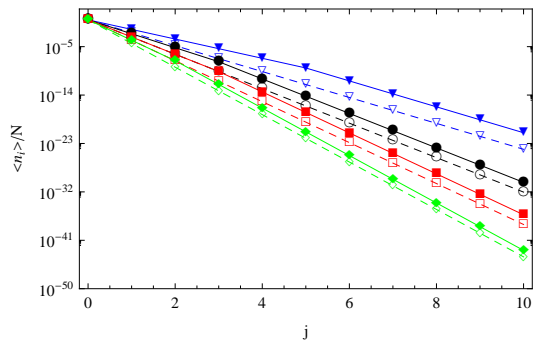}
\caption{(Color online) Atomic density distribution (in logarithmic scale) as a function of the distance from the center of the lattice for an open chain of $M=21$ sites and $N=21$ atoms. Solid lines and filled
symbols: Exact numerical solution of the variational problem. Dashed lines and empty symbols:
Approximate analytical solution $\chi = \Lambda / 8$. From top to bottom: Blue lines and triangles ($\Lambda=2.0$); black lines and circles ($\Lambda=5.0$); red lines and squares ($\Lambda=10.0$);
green lines and diamonds ($\lambda=20.0$). All quantities being plotted are dimensionless.}
\label{figteo}
\end{figure}
It is worth observing that the expression for $\chi$ that determines
the approximate solution to the maximally excited state of the system
does not depend on the size of the chain. On the other hand, it depends on the
number of atoms in the lattice through $\Lambda$.
In Fig.(\ref{figteo}) we have compared the exact
numerical solution of the variational problem with the approximate analytical solution for different
value of $\Lambda$ within the interval compatible with localization. One sees that even at moderate
values of $\Lambda$ the analytical approximation reproduces the essential features of the exact
localized quantum solution.

\begin{figure}[t]
{\includegraphics[width=7cm]{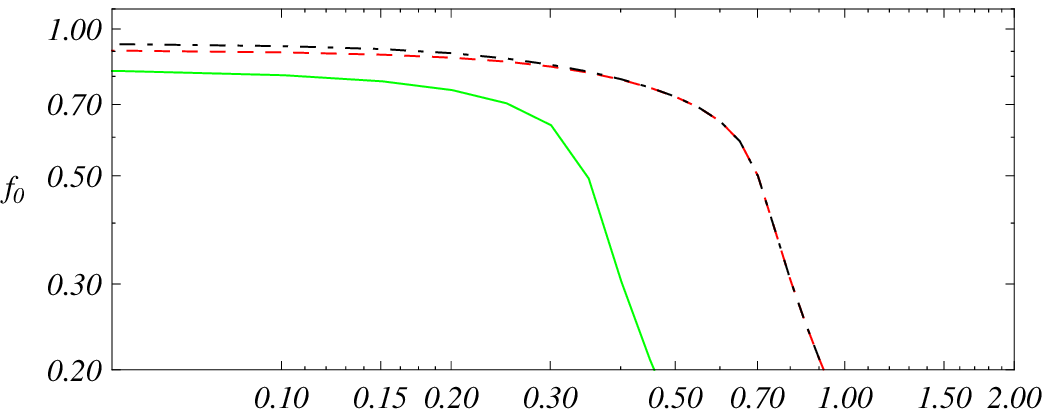}\vspace{-.19cm}
\includegraphics[width=7cm]{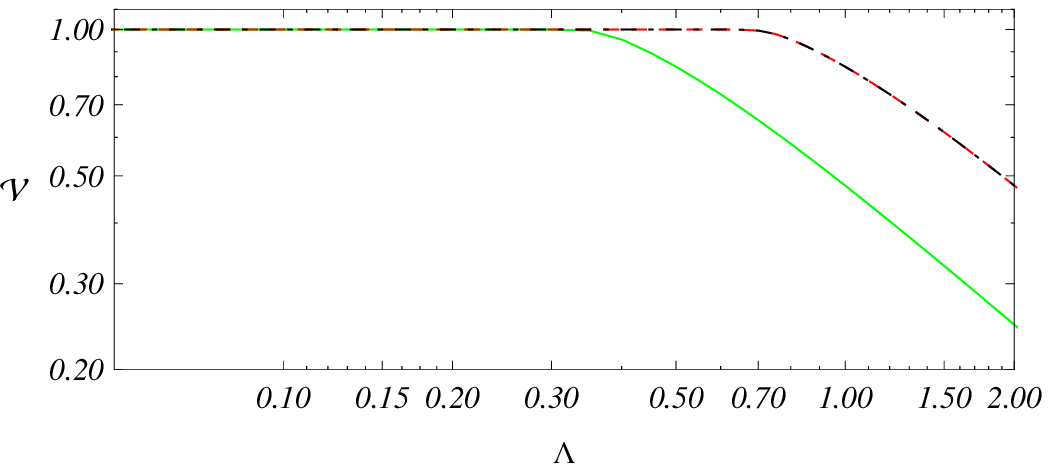}}
\caption{(Color online) Relative atomic population
ratio $f_0 = 1 - \langle n_0 \rangle /N$ (upper panel) and visibility
${\cal V}$ (lower panel) as functions of $\Lambda$ for open chains of different size
at unit filling. Green solid lines: Chain of $M=11$ sites and $N=11$ atoms.
Red dashed lines: Chain of $M=21$ sites and $N=21$ atoms.
Black dot-dashed lines: Chain of $M=31$ sites and $N=31$ atoms.
All quantities being plotted are dimensionless.}
\label{Visibility}
\end{figure}

The onset of localization in the maximally excited state can be naturally captured either by analyzing the behavior of the relative occupation of the central site $\langle n_0\rangle /N$ or by looking at the factor \mbox{$f_0 = 1 - \langle n_0\rangle /N$} that measures the relative atomic population outside the central site. The upper panel of Fig.(\ref{Visibility}) shows that for small values of $\Lambda$ the ratio $f_0$
is enhanced. In this regime and for very long chains, such that the border effects can be neglected,
$f_0 \rightarrow 1 - 1/M$. It begins to decrease at the onset of localization at the critical value $\Lambda_c \simeq 0.7$, finally vanishing asymptotically in the limit $\Lambda \rightarrow \infty$.

The rationale for the study of $f_0$ lies in the fact that it is directly associated
to the visibility $\cal V$, a quantity that can be actually measured by looking at
the relative difference between the maximum and the minimum of the momentum interference
pattern \cite{Firenze}. The visibility is related to the overall coherence \cite{Sengupta}
according to the relation
\begin{equation}
\label{eq1-sengupta}
{\cal V}=\frac{S_{max}-S_{min}}{S_{max}-S_{min}} \, ,
\end{equation}
where $S_{max}$ and $S_{min}$ are the maximum and minimum values of the momentum
distribution function:
\begin{eqnarray}
\label{eq2-sengupta}
S(K) &=& \frac{1}{M} \sum_{l,m=1}^M e^{i K (l-m)} \langle a^\dagger_l a_m \rangle
\end{eqnarray}
In the lower panel of Fig.(\ref{Visibility}) we report the behavior of the
visibility, as a function of $\Lambda$. We see that the visibility is very
sensitive to the onset of localization. It is approximately constant around its
maximum when the atoms are delocalized over the lattice at small
values of $\Lambda$ and begins to decrease exponentially, at the onset of the transition
in the center of the lattice, for $\Lambda_c \simeq 0.7$.
\begin{figure}[t]
\includegraphics[width=7cm]{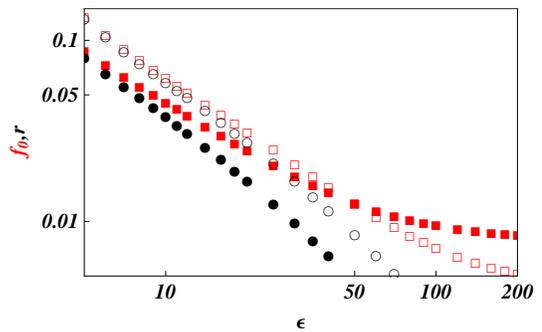}
\caption{(Color online) Relative atomic population outside the central site $f_0$ (red squares) and relative
population of the upper energy band $r$ (black circles) as a function of the gap width $\varepsilon$
for different values of the energy ratio $\Lambda$. Filled symbols: $\Lambda=2$.
Empty symbols: $\Lambda=3$. All quantities being plotted are dimensionless.}
\label{bandgap}
\end{figure}

\section{Role of higher bands}
A most serious issue concerns the
possibility that in the process of driving the system in the maximally excited state
of the lowest energy band, the lowest states of the higher energy bands may get significantly
populated. The problem is then to establish under what conditions the relative overlap between
these states and the maximally excited level of the lowest band is negligible. We thus need
to consider the two-band Bose-Hubbard Hamiltonian with intra-band and inter-band interaction
terms \cite{XuFisher}. One needs to add to the lowest-band Bose-Hubbard Hamiltonian
Eq.(\ref{BHH}) both the Bose-Hubbard Hamiltonian of the first upper energy band
\begin{equation}
\label{bcc2b-1}
H_2=\frac{U_2}{2} \sum^{d}_{ j=-d}  \hat n^{(2)}_j (\hat n^{(2)}_{j}-1) -
T_2 \sum^{d-1}_{j=-d}  \left (\hat a^{(2)\dagger}_{j} \hat a^{(2)}_{j+1} + \textrm{h.c.} \right ),
\end{equation}
and the inter-band interaction terms
\begin{equation}
\label{bcc2b-2}
H_I= E_g \sum^{d}_{ j=-d}  \hat n^{(2)}_j +
W \sum^{d}_{ j=-d} 4 \hat n_j \hat n^{(2)}_j+( \hat a^\dagger_{j}
\hat a^\dagger_{j} a^{(2)}_{j} \hat a^{(2)}_{j} + h.c.).
\end{equation}
In Eqs.(\ref{bcc2b-1},\ref{bcc2b-2}) $E_g$ is the energy gap between the first excited level of the
optical lattice potential and the relative ground state, while $\hat a^{(2)}_{j}$, $\hat a^{(2)\dagger}_{j}$, and $\hat n^{(2)}_{j}$ are, respectively, the on-site bosonic annihilation, creation, and number operators relative to the first upper energy band. The Hamiltonian parameters of the two bands are obviously not independent: resorting to the standard harmonic approximation for the optical lattice potential one has that $T_2 \simeq 9.4 T$,  $U_2 \simeq 3 U/4$, and $W \simeq U/2$. Hence, the total Hamiltonian
$H_T = H + H_2 + H_I$ depends only on the two independent parameters $\varepsilon=E_g/8T$,
that measures the width of the energy gap, and the previously introduced $\Lambda = U (N-1)/(8 T)$,
that expresses the ratio of the interaction to kinetic energy scales. Going again through the
same dynamical variational procedure for the two-band model $H_T$ and finite values of the gap
$\varepsilon$, we solve the maximization problem in the previously determined range of values
of $\Lambda$ that are compatible with exponential localization in the maximally excited state
of the lowest energy band. As we can see from Fig.(\ref{bandgap}), the relative atomic population
outside the central site $f_0$ is quite unaffected by the inter-band energy gap width until the latter becomes so small to allow a relative population $r= \sum_{j=-d}^d \langle n'_j \rangle/N$ of the first upper band that is comparable with $f_0$. If the energy gap is further reduced, the occupation outside the central site begins to increase exponentially even if the population of the central site remains a substantial fraction of the total number of atoms until $\varepsilon \simeq \Lambda$ and $E_g \simeq U (N-1)$.
This finding allows to conclude that if the lattice is loaded with a total number of atoms $N \le E_g/(10 U)$, one can safely disregard the presence of the upper energy bands.

\section{Conclusions and outlook}
In conclusion, we have introduced and discussed a mechanism of exponential localization in
the maximally excited state for systems of ultracold bosonic atoms with repulsive interactions
in open optical lattices. The properties of the maximally excited state have been studied
as a function of the Hamiltonian parameter both with numerical and analytical techniques
in order to determine the region of the parameter space in which exponential localization take place
and the dependence of the exponential decay on the Hamiltonian parameters. Finally, we have
discussed how the transition to localization can be detected experimentally by visibility
measurements, and we have established the physical conditions under which the overlap with
the upper energy bands can be neglected. This localization mechanism depends on the properties of the maximally excited state that stem from the interplay of nonlinearity and a bounded energy spectrum.
Being not a ground state property, it does not require the presence of random disorder,
attractive local potentials, or {\it ad hoc} truncations of the Hilbert space.

At first sight it would seem that to populate the maximally excited state of a system in realistic
experiments is an extremely challenging goal to achieve. However, thanks to the exceptional
properties of controllability and manipulability of optical lattice systems, the predicted
phenomenon might be observed first by cooling the system in the presence of a strong local
field favoring a substantial atomic population at the center of the lattice.
After switching off instantaneously the local field the system would remain,
with probability close to unity, in the strongly localized maximally excited state.
Repeated transitions to delocalization and exponential re-localization could then be
observed simply varying the energy ratio $\Lambda$ by changing the depth of the lattice
and/or by tuning the scattering length.

\acknowledgments

This work has been realized in the framework of the
FP7 STREP Project HIP (Hybrid Information Processing), Grant number
221889. The authors wish to thank Prof. Vittorio Penna for interesting and
stimulating discussions.

\vfill

\end{document}